\documentclass[aps,%
reprint,
prl,
groupedaddress]{revtex4-2}
\usepackage{color}
\usepackage{hyperref}
\usepackage{float}
\usepackage{amsmath,amsfonts,amssymb,epsfig,graphicx}
\usepackage{slashed}

\newcommand{\MGMCatNLO}{MadGraph5\_aMC@NLO}

\newcommand{\pt}{\ensuremath{p_\mathrm{T}}}

\newcommand{\fbinv}{\mbox{\ensuremath{~\mathrm{fb^{-1}}}}}

\newcommand{\al}{{\it et al. }}

\graphicspath{{./pic/}}

\begin{document}


\title{Boosted tau lepton as a microscope and macroscope}


\author{Sitian \surname{Qian}}
\email[]{stqian@pku.edu.cn}

\author{Zhe \surname{Guan}}
\email[]{zhe.guan@cern.ch}

\author{Sen \surname{Deng}}

\author{Yunxuan \surname{Song}}

\author{Tianyu \surname{Mu}}

\author{Jie \surname{Xiao}}

\author{Tianyi \surname{Yang}}

\author{Siguang \surname{Wang}}

\author{Yajun \surname{Mao}}

\author{Qiang \surname{Li}}
\email[]{qliphy0@pku.edu.cn}

\affiliation{State Key Laboratory of Nuclear Physics and Technology, School of Physics, Peking University, Beijing, 100871, China}

\author{Meng \surname{Lu}}
\email[]{meng.lu@cern.ch}

\author{Zhengyun \surname{You}}
\email[]{youzhy5@mail.sysu.edu.cn}

\affiliation{School of Physics, Sun Yat-Sen University, Guangzhou 510275, China}

\begin{abstract}
Anomalies from the LHCb lepton flavour universality and Fermilab muon anomalous magnetic moment, show tantalizing hints of possible new physics from the lepton sectors. Due to its large mass and shorter lifetime than muon, the tau lepton is believed to couple more to possible new physics beyond the Standard Model. Traditionally, tau leptons are probed  through their decay products due to tau's short lifetime. On the other hand, at a high energy scale, a large fraction of tau leptons could be boosted to a much longer life time and fly a visible distance from several centimetres up to kilometer length scale, yet very informative to new physics beyond the standard model or high energy cosmic rays. In this article, we investigate rare, yet promising, tau-physics phenomena, where long-lived taus are exploited either as a microscope (for the measurement of tau's anomalous magnetic moment to an unprecedented level of accuracy) or as a  macroscope (for the detection of 1 TeV to 1 PeV cosmic neutrinos).
\end{abstract}

\maketitle

\section{Introduction.}
\label{introduction}

The LHCb Collaboration recently reported an update on testing lepton-flavour universality with $B^+ \rightarrow K^+ \ell^+ \ell^-$, in which 3.1 standard deviations from the Standard Model (SM) prediction were observed~\cite{Rk1}. The Muon g-2 Experiment at Fermilab also has  published its new results, leading to a new world average of mismatch between theory and experiment at 4.2 standard deviations~\cite{gminus2}. These discrepancies show interesting tension with the SM in the muon sector.

Compared with the muon, the tau has a larger mass (around 1.777 GeV), thus it is believed to be more sensitive to new physics beyond the SM (BSM)~\cite{Dam:2018rfz}. However, the mean life time of the tau-lepton is around $2.9\times 10^{-13}$ s~\cite{RPP}, and the mean flying distance is correspondingly at  87 $\mu$m, making the direct detection of the tau-lepton very difficult at the current experiments

Normally, tau leptons are probed through their decay products. But if the energy is around or above 1 TeV, a large fraction of tau leptons could be boosted to a much longer life time and fly a visible distance: the decay length of $\langle L_{\tau} \rangle \sim 5$~cm ($50$~m) $\cdot \, E_{\tau}$ / TeV (PeV), where $ E_{\tau}$ is the tau energy. This brings great potential to exploit highly boosted taus, which should have already been produced at current LHC experiments or from the cosmic ray and will be produced copiously at many future facilities.
 
In the next decades, the LHC and the High-Luminosity LHC (HL-LHC), together with other future colliders in design, will be further exploring the SM and searching for physics beyond that. The majority of the proposed future machines are lepton colliders, designed primarily for Higgs-boson measurements. The most promising proposals include linear or circular electron-positron colliders~\cite{ILC, FCC, CEPC, CLIC} and muon colliders~\cite{Muc0,Muc1,Muc2,Muc3,Muc4}.  

At the LHC and the high energy electron or muon colliders, tau leptons can be produced through decays of produced particles (mostly W/Z bosons and heavy quarks) to $\tau\nu$, $\tau\tau$, $\tau\tau\gamma$ or $\gamma\gamma\rightarrow \tau\tau$ process, etc. With boosted taus, one has the possibility to directly observe tau lepton and measure its anomalous magnetic moment $a_\tau = \frac{1}{2}(g_\tau-2)$ to an unprecedented level of accuracy, as will be shown below. 

Very high energy tau leptons can also be produced from primary cosmic rays interacting with the earth or atmosphere~\cite{Reno:2019jtr}. For example, the IceCube experiment reported observations of high-energy astrophysical neutrinos, with several tens neutrino candidate events with deposited energies ranging from 30 TeV to 2 PeV~\cite{IceCube:2014stg}, together with two tau neutrino candidate events with deposited energy of about 100 TeV and 2 PeV~\cite{IceCube:2020abv}. The boosted taus arising from the inverse beta decay process, can fly to a distance up to hundreds of meters, and a compact underground experiment with tracker and calorimeter detectors should be able to capture them, to probe the source of mysterious cosmic rays and to study the Universe's most powerful cosmic accelerators~\cite{IceCube:2018cha}.

\section{Long flying tau and the benefit on reconstruction}
\label{llt}
Due to its short life time, tau was observed and measured previously mostly indirectly through its decay products in experiments, apart from Chorus~\cite{CHORUS:1998gsq} and DONUT ~\cite{DONUT:2000fbd}, which are emulsion-based experiments. At high energy, tau can be boosted to a much larger life time and fly a more visible distance. A real tau track and its decay vertex can essentially be reconstructed with hits in current silicon and pixel detectors, at e.g. the LHC and future collider experiments, giving the possibility to ``see" tau directly. 

The tau track consists of several hits, and can be matched with a displaced vertex in the 3-prong case which can be reconstructed from the tracks of tau's decay products. While in the 1-prong and leptonic decay cases, the tau track can be combined with the emerging track (to form a kink) from tau's decay product to reconstruct displaced vertex (otherwise, only impact parameter can be derived if there are no tau hits and thus no constraint available for the global track fitting).

\begin{figure}
    \centering
    \includegraphics[width=0.5\textwidth]{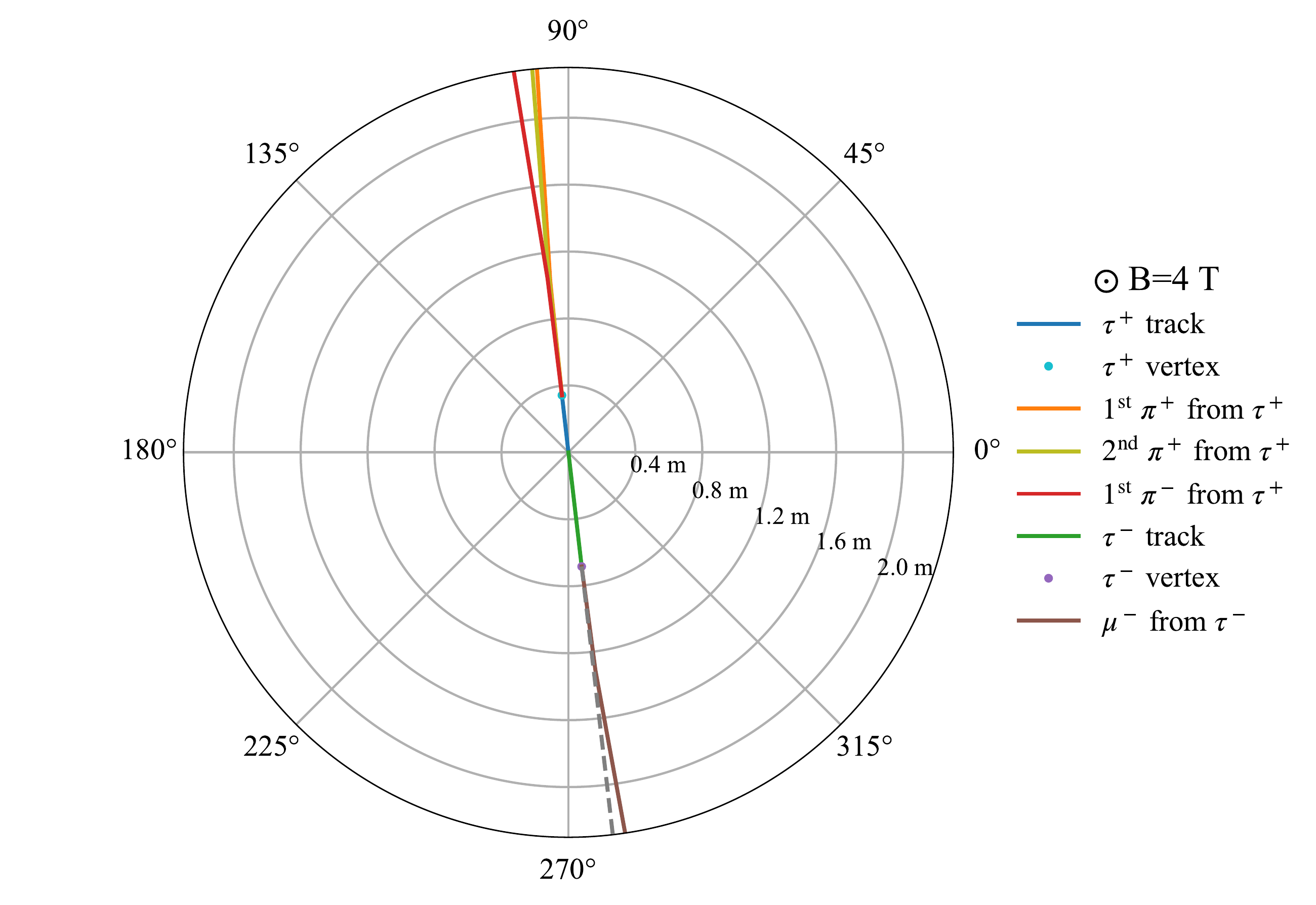}
    \caption{Sample tracks from long lived $\tau$ lepton produced at a muon collider with center-of-mass energy equals 3 TeV. $\tau^+$ undergoes 3-prong decay while $\tau^-$ decays to $\mu^-$ plus two neutrinos. For illustration purposes, $\tau^+$($\tau^-$) lifetime is set to 5(10) times the mean life time. The blue and green solid lines indicate the tau tracks which can be matched with the displaced vertex (the cross point of the three charged pions) and the decayed muon track (in purple), respectively. The green dashed line indicates the original direction of muon velocity.}
    \label{fig:3prong}
\end{figure}

Let us take an example, tau pair production event with one decay in 3-prong and the other 1-prong. The decay pattern can be plotted in the cartoon plot, fig.~\ref{fig:3prong}, which shows a typical example event for $\tau^+\tau^-$ productions at a 3 TeV muon collider, where both tau leptons have transverse momentum $\pt=1.4$\,TeV, and pseudorapidity $|\eta|=0.37$. For illustration purposes, $\tau^+$ ($\tau^-$) lifetime is set to 5 (10) times of the mean lifetime. The $\tau^+$ undergoes a 3-prong decay of $\pt(\pi^-)=297$\,GeV, and the other two $\pi^+$ have $\pt$ as 662 and 271 GeV, respectively,  while $\tau^-$ decays to $\mu^-$ plus two neutrinos, with muon $\pt=160$\,GeV. The separation distance among decay products is $\Delta\eta\times\Delta\phi\sim 0.002\times 0.002$. However, the decayed products are with modest momentum, and the displaced tau decay vertex can be reconstructed with reasonable efficiency around 80\%, as e.g. shown by the CMS long-lived particle searches~\cite{CMS:2021tkn}. The performance can be further improved with kinematic fit and machine learning techniques~\cite{Cardini:2021bar}. For example, the identification efficiency and mis-identification rate of boosted tau at the CMS experiment, are at 80-90\% and below $10^{-3}$, respectively~\cite{CMS:2018jrd,DP2016038}. 

For 5 TeV tau leptons produced at a 10 TeV muon collider through $\mu\mu\rightarrow\tau^+\tau^-$, one can estimate the fraction of taus flying a distance larger than 10 cm to be 32\%, based on the exponential decay formula. The signatures include at least one long-lived tau-tracks matched with 1-prong or 3-prong trackers from tau decays, together with photons and missing energy, and should be able to be reconstructed with high efficiency and reasonable resolution as mentioned above. The event can be triggered by requiring a displaced vertex of distance larger than 5 cm  from the prompt interaction point~\cite{CMS:2021tkn}. Considering all the above factors, i.e. tau life time probability, displaced vertex tagging efficiency and tau decay branch ratios, we estimate the final event selection efficiency with at least one long-haul flying tau to be around 20--40\%.

\section{Anomalous Magnetic Moment}
\label{amm}
Precise measurement of the anomalous magnetic moment (g-2) and Electric Dipole Moment (EDM) of particles are important tests of physics BSM. It is generally believed that the tau lepton is more sensitive to BSM due to its larger mass, while on the other hand, tau's EDM and g-2 can only be searched for at collider experiments due to their short life time. 

The theoretical prediction on anomalous magnetic moment $a_{\tau,\, \text{SM}}^\text{pred}$ is at $10^{-5}$ precision~\cite{Eidelman:2007sb}. The most precise probe on the experimental value ($a_\tau^\text{exp}$) is from DELPHI~\cite{RPP,Abdallah:2003xd} at the Large Electron Positron Collider (LEP), but is an order of magnitude higher than the theoretical central value of $a_{\tau,\, \text{SM}}^\text{pred}$.
\begin{align}
    a_\tau^\text{exp} =  - 0.018\,(17),\quad a_{\tau,\,\text{SM}}^\text{pred} = -0.001\,177\,21\,(5).
\end{align}

Various methods have been proposed, aiming at improving the sensitivity of measuring $a_\tau$ at the current and future facilities. For example, the precision can be improved by one order of magnitude beyond DELPHI~\cite{Abdallah:2003xd}, with the HL-LHC, through heavy Ion collisions~\cite{Beresford:2019gww}, or even more at superKEKB~\cite{Crivellin:2021spu}. At the proposed 3-TeV Compact Linear Collider (CLIC), the precision is estimated to be able to reach $|a_\tau|\approx 3\times 10^{-3}$ with 2000\fbinv of data, through the $\gamma\gamma$ and $e\gamma$ collisions~\cite{Ozguven:2016rst}. Studies have also been performed at the HL-LHC through $H\gamma$ productions and the Higgs rare decay into $\tau\tau\gamma$~\cite{Cao:2021trr}, where the projected sensitivity is evaluated in the framework of effective field theory and can extend the LEP searches to a large extent. Similar studies have also been done at the muon colliders~\cite{Koksal:2018vtt,Buttazzo:2020eyl}.

We use SM effective field theory (EFT)~\cite{Escribano:1996wp} to introduce the BSM modifications of $a_\tau$ and $d_\tau$, similarly as done in Ref.~\cite{Beresford:2019gww}. Two dimension-six operators in the Warsaw basis~\cite{Grzadkowski:2010es} modify $a_\tau$ and $d_\tau$ at tree level, as discussed in ~\cite{Beresford:2019gww}:
\begin{align}
    \mathcal{L}' =  \left(\bar{L}_{\tau}\sigma^{\mu\nu} \tau_R\right) H
    \left[
    \frac{C_{\tau B}}{\Lambda^2} B_{\mu\nu} +
    \frac{C_{\tau W}}{\Lambda^2} W_{\mu\nu}   \right],
    \label{eq:BSMLagrangian}
\end{align}
where $\tau_R$ is the spinor with R denoting chirality, $\Lambda$ represents the BSM physics scale, $C_i$ are complex dimensionless Wilson coefficients, $B_{\mu\nu}$ and $W_{\mu\nu}$ are the U(1)$_\text{Y}$ and SU(2)$_\text{L}$ field strengths, and $H$ ($L_\tau$) is the Higgs (tau lepton) doublet. We fix $C_{\tau W} = 0$ to parameterize the two modified moments $(\delta a_\tau, \delta d_\tau)$ using two real parameters $(|C_{\tau B}|/\Lambda^2, \varphi)$~\cite{Eidelman:2016aih}
\begin{align}
    \delta a_\tau &= \frac{2m_\tau}{e}\frac{|C_{\tau B}|}{M}\cos \varphi,\quad
    \delta d_\tau = \frac{|C_{\tau B}|}{M} \sin \varphi,
    \label{eq:delta_a_d_tau_defn}
\end{align}
where $v = 246$~GeV, the SM Higgs vacuum expectation value, $\theta_W$ is the Weinberg angle, $\varphi$ is the complex phase of $C_{\tau B}$, and $M$ is defined to be $\Lambda^2/ (\sqrt{2}v\cos\theta_W)$. 

In this study, we are interested in highly boosted tau leptons at the TeV scale, which offers a two-fold bonus. Firstly, the energetic tau can be very sensitive to the BSM modifications of $a_\tau$ and $d_\tau$. Secondly, a large fraction of boosted tau can fly a long distance, providing the possibility to fully reconstruct the tau track together with its decay pattern as discussed above.

We focus on the production process of $\mu\mu\rightarrow \tau\tau\gamma$ (with highly energetic tau) and $\gamma\gamma\rightarrow \tau\tau$ (with mildly energetic tau) at a 10 TeV muon collider. Checks have also been done at a 3 TeV muon collider. Both signal and background (the SM predictions with EFT parameters set as 0) events are simulated with \MGMCatNLO. At generator level, $\pt>5$ GeV and $|\eta|<3$ are applied for photons and leptons, together with $\Delta R_{l\gamma}>0.1$ (where $\Delta R=\sqrt{((\Delta\Phi)^2 +\Delta\eta)^2 }$ and 
$\Delta\Phi$ and $\Delta\eta$ the differences of azimuthal angle and pseudorapidity between lepton and photon).  Equivalent photon approximation is adopted to simulate $\gamma\gamma\rightarrow \tau\tau$ in \MGMCatNLO.  

\begin{figure}
    \centering
    \includegraphics[width=0.4\textwidth]{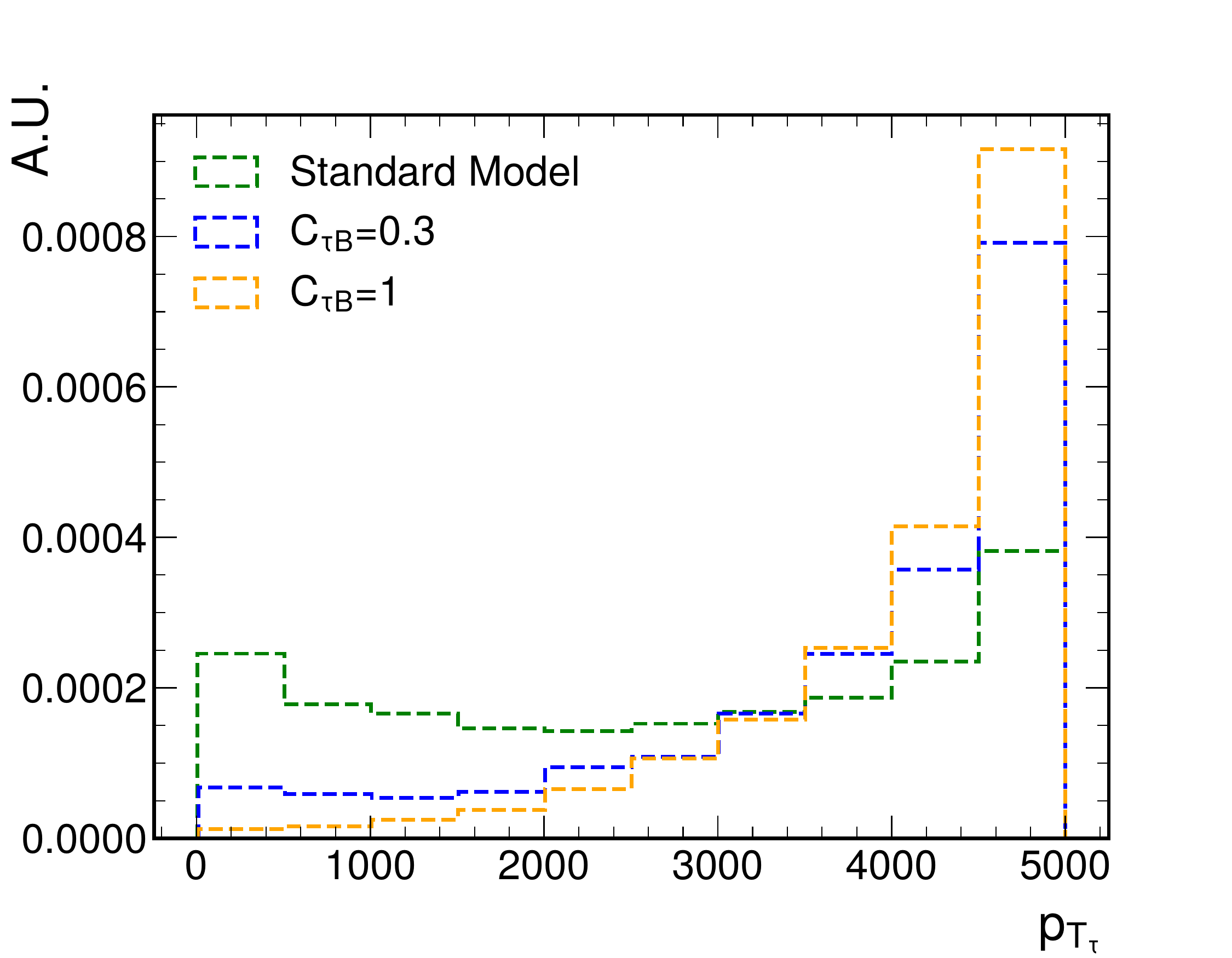}
    \caption{ $\pt$ of $\tau$ distributions and the dependence on EFT parameters for $\mu\mu\rightarrow \tau\tau\gamma$ productions at a 10 TeV muon collider.}
    \label{fig:taupt}
\end{figure}

\begin{figure}
    \centering
    \includegraphics[width=0.4\textwidth]{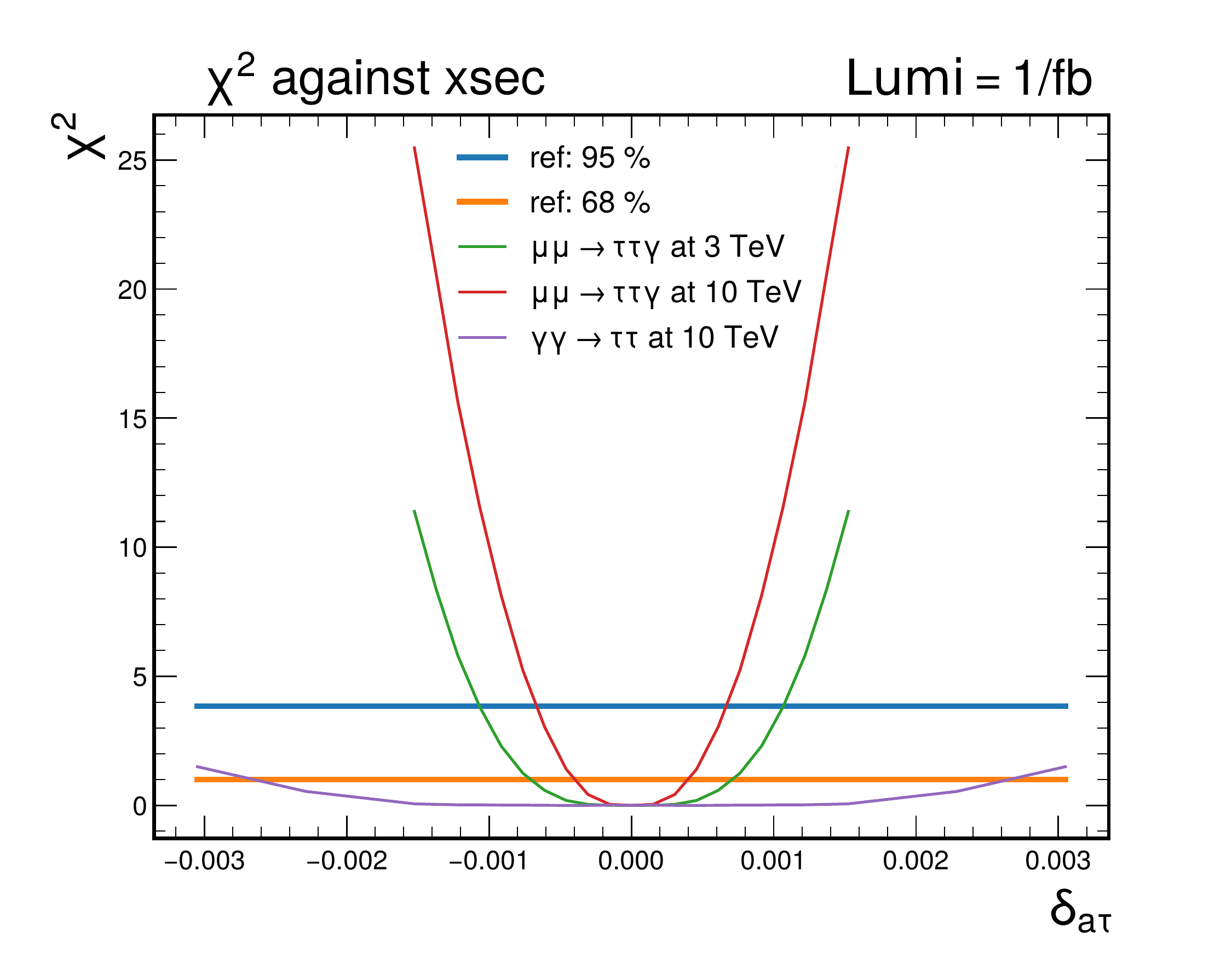}
    \caption{Projected sensitivity on tau g-2 for both $\mu\mu\rightarrow \tau\tau\gamma$ and $\gamma\gamma\rightarrow \tau\tau$ processes, at a 3 or 10 TeV muon collider with 1 \fbinv and perfect event efficiency is assumed.}
    \label{fig:limit}
\end{figure}

Fig.~\ref{fig:taupt} shows the $\pt$ of $\tau$ distributions and the dependence for three values of the $C_{\tau B}$ EFT parameter for $\mu\mu\rightarrow \tau\tau\gamma$ process at a 10 TeV muon collider. One can see $\mu\mu\rightarrow \tau\tau\gamma$ process is indeed very sensitive to tau g-2 related EFT parameters. Based on generator level events, we then estimate the sensitivity by scanning $a_{\tau}$ and counting events number. The sensitivity results are shown in Fig. ~\ref{fig:limit}.  One can see with only 1 \fbinv of integrated luminosity, $\mu\mu\rightarrow \tau\tau\gamma$ can already be sensitive to tau g-2 at an accuracy of $10^{-3}$ ($10^{-4}$) at the 95\% confidence level, surpassing much over the sensitivity from $\gamma\gamma\rightarrow \tau\tau$. However, when taking into account the event selection efficiency as discussed above, 10-100 \fbinv is needed to reach this accuracy.

\section{Astrophysical Tau neutrino}
\label{lhcastro}

Primary cosmic rays (protons or neutrinos) can interact with the earth, its surrounding atmosphere or materials in the detectors, and convert to secondary particles, including highly boosted tau~\cite{Reno:2019jtr}. These energetic taus can be used to probe the sources of cosmic rays and the mechanism of cosmic acceleration, and will provide valuable information on flavor composition of cosmic neutrinos. For example, IceCube probed the astrophysical flux down to 35 TeV and showed consistency with the $(f_e:f_{\mu}:f_\tau)_\oplus\approx(1:1:1)_\oplus$ flavor ratio at Earth, as commonly expected from the averaged oscillations of neutrinos produced by pion decay in distant astrophysical sources~\cite{IceCube:2015rro}. 

In the last decades, experiments including Super-Kamiokande~\cite{Super-Kamiokande:2012xtd} and IceCube~\cite{IceCube:2019dqi} have measured atmosphere tau neutrinos with energy up to 50-60 GeV. At or above 10-100 TeV scale, astrophysical neutrinos start dominating over atmosphere ones~\cite{IceCube:2014rwe,IceCube:2015rro}. For example, IceCube reported two tau neutrino candidate events from IceCube with deposited energy of about 100 TeV and 2 PeV~\cite{IceCube:2020abv}. Above EeV scale, ANITA~\cite{ANITA:2016vrp} reported two anomalous events which look like Extensive Air Shower from EeV scale tau neutrinos. There have been many proposals for future experiments focusing on searching for high energy tau neutrinos such as CRTNT~\cite{Cao:2004sd}, POEMMA~\cite{Venters:2019xwi}, TRINITY~\cite{Brown:2021lef} and GRAND~\cite{Fang:2017mhl}. 

The information on cosmic tau neutrino at the energy scale between 1 TeV and 100 TeV, however, is mostly missing. This is due to the fact that observing the resulting tau lepton from tau neutrino interaction is usually quite difficult, and distinguishing tau neutrino from the other lepton species is hard~\cite{Super-Kamiokande:2012xtd}. Moreover, at IceCube, double cascade events become identifiable only above deposited energies of 60 TeV, which is crucial to break the degeneracy between electron and tau neutrino flavors~\cite{IceCube:2020abv}. 

\begin{figure}
    \centering
    \includegraphics[width=0.4\textwidth]{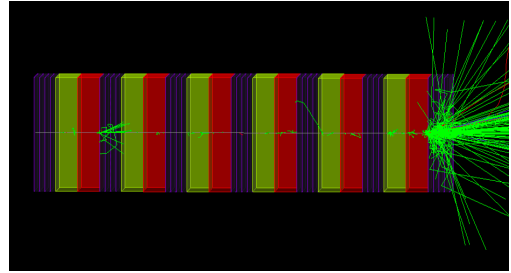}
    \caption{Geant4 simulation for a 50 TeV tau penetrating interleaved detectors made by 3 layer RPC (in blue), lead tungstate (PbWO4, in green) and copper (in red), each with 20 cm's length. The energetic tau flies like a muon and then decays into 3 charged pions and neutrinos. The gamma rays shows in green, electrons in red and electron neutrinos in white.}
    \label{fig:taudet}
\end{figure}

Here we propose a novel type of high energy tau neutrino experiment in an underground laboratory, aiming at tau neutrinos with energy at around 10 TeV to 1 PeV scale. The experiment is not necessary to be very deep underground, as our signals are quite energetic and thus, contamination from cosmic muons should be easy to be suppressed. The detector is cubic-like with a size of 50-100 meters, consisting of interleaved tracker (or nuclear emulsion) and calorimeter modules. The detector materials, together with the surrounding rocks, serve as the target to capture high energy neutrinos. Those tau leptons produced in nearby rock entering the detector can produce track ending with decay point signature as shown in Fig.~\ref{fig:taudet}, while those produced inside detectors can have a signature involving double cascades connected by a track~\cite{IceCube:2020abv}. This kind of underground experiment can be seen as the detector of the underground cosmic ray collider, sensitive to boosted taus with a Lorentz factor up to $10^{4-6}$.

Fig.~\ref{fig:taudet} shows such a detector modular as proposed above, where we simulated with Geant4~\cite{GEANT4:2002zbu} a 50 TeV tau penetrating interleaved detectors made by 3 layer Resistive Plate Chambers (RPC, in blue), lead tungstate (PbWO4, in green) and copper (in red), each with 20 cm's length (one can also insert these into layers of rocks). The energetic tau flies like a muon and then decays into 3 charged pions and neutrinos, depositing energy in calorimeters. With such a detector, one should be able to single out long flying taus decaying into electrons or hadrons. The track and decay length can provide direct information on the direction and energy scale of original cosmic neutrino, respectively. The energy deposit in detectors can be used to regress the energy of produced leptons and thus original neutrinos.

For tau neutrinos with energy larger than $E_0\sim$ 10 TeV (100 TeV) interacting with the material, the mean flying distance of produced tau leptons is around 0.5 (5) meter. With a 100-150 meter cubic detector as mentioned above, the surrounding earth plus detector materials can be seen as a hemisphere with a radius of roughly $R=100-200$ meter. The expected event yield can be roughly estimated with:
\begin{align} \label{nyield}
dN/dt &=\int^{\infty}_{E_0}2\pi\cdot\Phi(E)\cdot\sigma_{n\nu}\cdot \rho V N_A\cdot dE, \nonumber\\
&\sim 2\pi E_0\cdot\Phi(E_0)\cdot\sigma_{n\nu}\cdot \rho V N_A,
\end{align}
where $N_A$ is the Avogadro constant, $\rho=$ 5.514 $\rm{g}/\rm{cm}^3$ and $V=R^3$ (should be larger if taking into account nearby earth materials) are the mean density and volume of the target material, $\sigma_{n\nu}$ symbols the neutrino nucleon cross sections and is around $10^{-34} (10^{-33})\rm{cm}^2$ for a 10 TeV (100 TeV)  neutrino~\cite{IceCube:2020rnc,Formaggio:2012cpf}, $\Phi(E)$ is the high energy neutrino tau flux which at 10 TeV (100 TeV) scale can be estimated to be around $10^{-7/-8}/E^2\,\rm{GeV}\,\rm{cm}^{-2}\rm{s}^{-1}\rm{sr}^{-1}$~\cite{IceCube:2020abv,IceCube:2020wum,IceCube:2021uhz}. 
Taking into account all these together, one can expect ${\cal O} [10\, (1)]$ events for a 5 years run (for $E_0\sim$ 5 TeV, the event yields can be further larger by a factor about 4 or 5). The estimation is without detector efficiency though.



\begin{figure}
    \centering
    \includegraphics[width=0.36\textwidth]{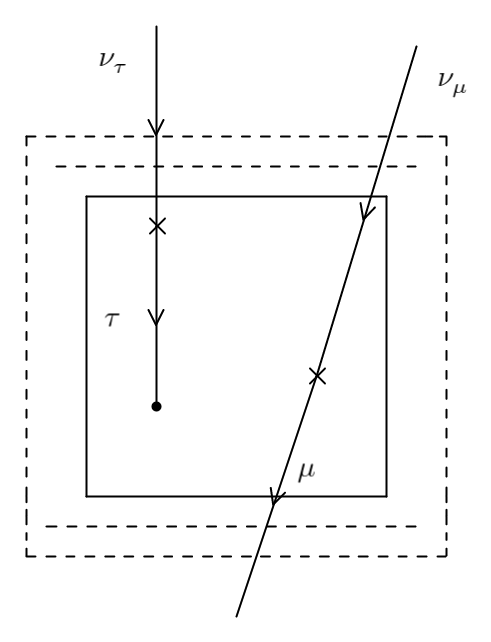}
    \caption{An illustration of the proposed underground experiment to catch cosmic tau and muon neutrinos with energy at 10 TeV and above. The 100-meter size cube in the center represents the tracker and calorimeter modules as shown in Fig.~\ref{fig:taudet}. The dashed lines are for RPC detectors. The surrounding rocks can effectively enhance the target volume on top of the detector's materials. Two example events are shown in the plot for $\nu_\mu \rightarrow \mu$ passing through the detector leaving a long track, and $\nu_\tau \rightarrow \tau$ with $\tau$ decay in the cube. The cross points symbol the interactions points where neutrinos turn into lepton. The dot point symbols the place where tau lepton decays.}
    \label{fig:taudet2}
\end{figure}

Fig.~\ref{fig:taudet2} is an illustration of the proposed underground experiment to catch cosmic tau and muon neutrinos with energy at around the 10 TeV scale. The 100-meter size cube in the center represents the tracker calorimeter modules as shown in Fig.~\ref{fig:taudet}. The dashed lines are for RPC detectors. 
Two example events are shown in the plot for $\nu_\mu \rightarrow \mu$ passing through the detector leaving a long track, and $\nu_\tau \rightarrow \tau$ with $\tau$ decay in the cube. Note a TeV scale muon will fly a distance typically around 6000 km so unlikely to decay within the detector.

The overall detector can be made modular or in grid shape, and can be expanded gradually in the underground hall or tunnel. Such an experiment should be able to measure the direction of cosmic neutrino sources accurately and separate tau neutrino from electron/muon neutrinos, and provide energy information through the tau flying distance and the calorimeter deposits. Furthermore, it may also be sensitive to earth skimming neutrino at EeV scale or above.

\section{Conclusions}
\label{discussions}

We propose unique physics can be done with energetic tau leptons at or above TeV scale, which can be produced from colliders or cosmic accelerators. A large fraction of these tau leptons could be boosted to a much longer life time and fly a visible distance from several centimetres to hundred meters. Matching tau track hits with displaced vertex reconstructed from the tau decay products, one should be able to observe long flying tau leptons with high efficiency and accuracy. We study rare yet promising tau physics by exploiting those long-lived taus, including, measuring its anomalous magnetic moment to an unprecedented level of accuracy, and detecting high energy cosmic tau neutrinos at the 1 TeV to 1 PeV scale. Similar studies can also be applied to, e.g. long-lived particle searches, and charged lepton flavor violation through $\mu\mu\rightarrow\tau\mu$. Performance can be further improved with dedicated detector R\&D (e.g. a high granularity calorimeter, and/or a timing detector).

\appendix

\begin{acknowledgments}
This work is supported in part by the National Natural Science Foundation of China under Grants No. 12150005, No. 12075004 and No. 12061141002, by MOST under grant No. 2018YFA0403900. A preprint has previously been published~\cite{Qian:2022owu}.
\end{acknowledgments}


\bibliographystyle{ieeetr}
\bibliography{h}
\end{document}